\documentclass[prb,superscriptaddress,twocolumn,showpacs,preprintnumbers,amsmath,amssymb]{revtex4}

\usepackage{graphicx}
\usepackage{dcolumn}
\usepackage{bm}

\begin{document}

\title{Weak localization scattering lengths in epitaxial, and CVD graphene}
\date{\today}
\author{A.M.R. Baker}
\affiliation{Dept. of
Physics, University of Oxford, Clarendon Laboratory, Parks Rd.,
Oxford, OX1 3PU, U.K.}

\author{ J.A. Alexander-Webber}
\affiliation{Dept. of
Physics, University of Oxford, Clarendon Laboratory, Parks Rd.,
Oxford, OX1 3PU, U.K.}

\author{T. Altebaeumer}
\affiliation{Dept. of
Physics, University of Oxford, Clarendon Laboratory, Parks Rd.,
Oxford, OX1 3PU, U.K.}


\author{T.J.B.M. Janssen}
\affiliation{National Physical Laboratory, Hampton Road, Teddington, TW11 0LW, U.K.}

\author{A. Tzalenchuk}
\affiliation{National Physical Laboratory, Hampton Road, Teddington, TW11 0LW, U.K.}

\author{S. Lara-Avila}
\affiliation{Dept. of Microtechnology and Nanoscience, Chalmers University of Technology, S-412 96 G$\ddot{o}$teborg, Sweden}

\author{S. Kubatkin}
\affiliation{Dept. of Microtechnology and Nanoscience, Chalmers University of Technology, S-412 96 G$\ddot{o}$teborg, Sweden}

\author{R. Yakimova}
\affiliation{Dept. of Physics, Chemistry and Biology, Link$\ddot{o}$ping University, S-581 83 Link$\ddot{o}$ping, Sweden}

\author{C.-T. Lin}
\affiliation{Institute of Atomic and Molecular Sciences, Academia Sinica, Taipei, 11617, Taiwan}

\author{L.-J. Li}
\affiliation{Institute of Atomic and Molecular Sciences, Academia Sinica, Taipei, 11617, Taiwan}

\author{R.J. Nicholas}
\email{r.nicholas1@physics.ox.ac.uk}
\affiliation{Dept. of
Physics, University of Oxford, Clarendon Laboratory, Parks Rd.,
Oxford, OX1 3PU, U.K.}
\begin{abstract}

Weak localization in graphene is studied as a function of carrier density in the range from 1 x $10^{11}$\,cm$^{-2}$ to 1.43 x $10^{13}$\,cm$^{-2}$ using devices produced by epitaxial growth onto SiC and CVD growth on thin metal film. The magnetic field dependent weak localization is found to be well fitted by theory, which is then used to analyse the dependence of the scattering lengths L$_\varphi$, L$_i$, and L$_*$ on carrier density. We find no significant carrier dependence for L$_\varphi$, a weak decrease for L$_i$ with increasing carrier density just beyond a large standard error, and a n$^{-\frac{1}{4}}$ dependence for L$_*$. We demonstrate that currents as low as 0.01\,nA are required in smaller devices to avoid hot-electron artefacts in measurements of the quantum corrections to conductivity.


\end{abstract}


\maketitle

\section{Introduction}

In recent years graphene has proved of great interest both for its huge range of potential applications, from enhancing the strength of composite materials\cite{Stankovich2006}, to high-speed analogue electronics\cite{Bourzac2012}; and for its impressive range of physical properties, including an anomalous integer quantum Hall effect\cite{Novoselov2005}, quantized opacity\cite{Nair2008}, and its two-dimensionality\cite{Novoselov2005}. Amongst other properties it shows a greatly enhanced weak (anti)localization effect\cite{Novoselov2005}, which is the principal topic of this paper.

The nature of weak (anti)localization in graphene has attracted a significant amount of controversy\cite{Geim2007}. It was originally predicted that the effect would be entirely of the weak antilocalization type due to the existence of a Berry phase in graphene. Early results, however, failed to show such behaviour\cite{Geim2007}. Subsequently, it was realized that this could be resolved by the addition of further scattering terms which break chirality, particularly elastic intervalley scattering\cite{McCann2006}.

The purpose of this paper is the fitting of scattering lengths using the theory of McCann et al.\cite{McCann2006} for a wide range of different graphene samples. The fittings are used to demonstrate the validity of this method for devices with carrier densities ranging from 1 x $10^{11}$\,cm$^{-2}$ to 1.43 x $10^{13}$\,cm$^{-2}$. Devices are analysed from graphene produced by epitaxial growth on SiC\cite{Janssen2011}, and chemical vapour deposition (CVD) onto thin metal films\cite{Chen2012}. The results are compared with those obtained from the literature\cite{Lara2011b,Tikh2008,Ki2008,Jaur2011} and together are used to measure trends in the scattering lengths with carrier density.

We also demonstrate that measurements of the the dephasing length at the lowest temperatures can be significantly influenced by hot electron effects\cite{Baker2012,Baker2012b}. The currents required to avoid this effect are calculated and are demonstrated to be as low as 0.01\,nA for small devices.

\section{Methodology and Theoretical Background}

Hall bar devices were produced using graphene derived from the epitaxial and CVD fabrication methods. The devices were produced using e-beam lithography and oxygen plasma etching. The epitaxial graphene was grown on the Si-terminated face of SiC\cite{Janssen2011}, with contacts made using large area titanium-gold contacting. Photochemical gating was used to control the carrier density on the epitaxial devices due to the impossibility of conventional backgating through SiC\cite{Lara2011}. CVD graphene was grown on thin-film copper, subsequently transferred to Si/SiO$_2$, and contacts were made using chrome-gold tracks/bondpads followed by gold-only final contacting, as described in our previous work\cite{Baker2012}. Various sizes of large-area Hall bar were produced; dimensions were typically 64 x 16\,$\mu$m$^2$ for the CVD devices, and 160 x 35\,$\mu$m$^2$ for the epitaxial devices. Considerable care was taken to record the magnetotransport data with the use of slow magnetic field sweep rates passing completely through the zero field resistance peak. Measurements of the phase of Shubnikov-deHaas and Quantum Hall effect oscillations at higher fields\cite{Baker2012b} demonstrate that all samples studied were monolayer graphene with charge density fluctuations less than the measured carrier density.

\begin{figure}
\begin{center}
\includegraphics[width = 8cm]{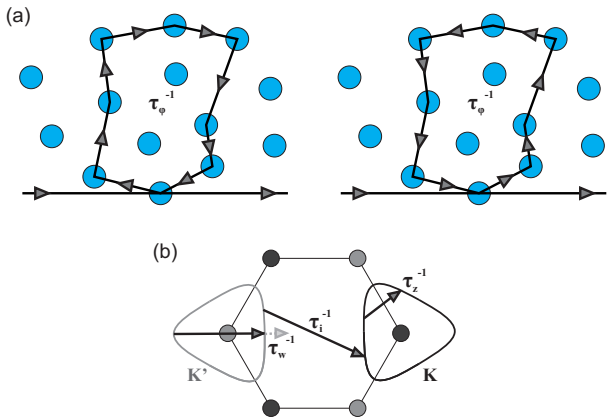}
\end{center}
\caption{Illustration of the scattering processes which contribute to weak (anti)localization. (a) Two example scattering paths, identical except for the direction of travel around the loop. The dephasing rate, $\tau^{-1}_\varphi$, controls the maximum size of such loops due to the need for phase coherence to produce an interference effect. (b) The two rounded triangles centred on the two inequivalent Dirac points, K, K' are shown for a small Fermi energy such that trigonal warping is clearly apparent. Three scattering terms, and how they contribute, are superimposed on this Fermi surface: $\tau^{-1}_i$, the elastic intervalley scattering rate; $\tau^{-1}_w$, the elastic intravalley trigonal warping scattering term and $\tau^{-1}_z$ the elastic intravalley chirality breaking scattering term.} \label{fig:wltaus}
\end{figure} 

\begin{figure*}
\begin{center}
\includegraphics[width = 18cm]{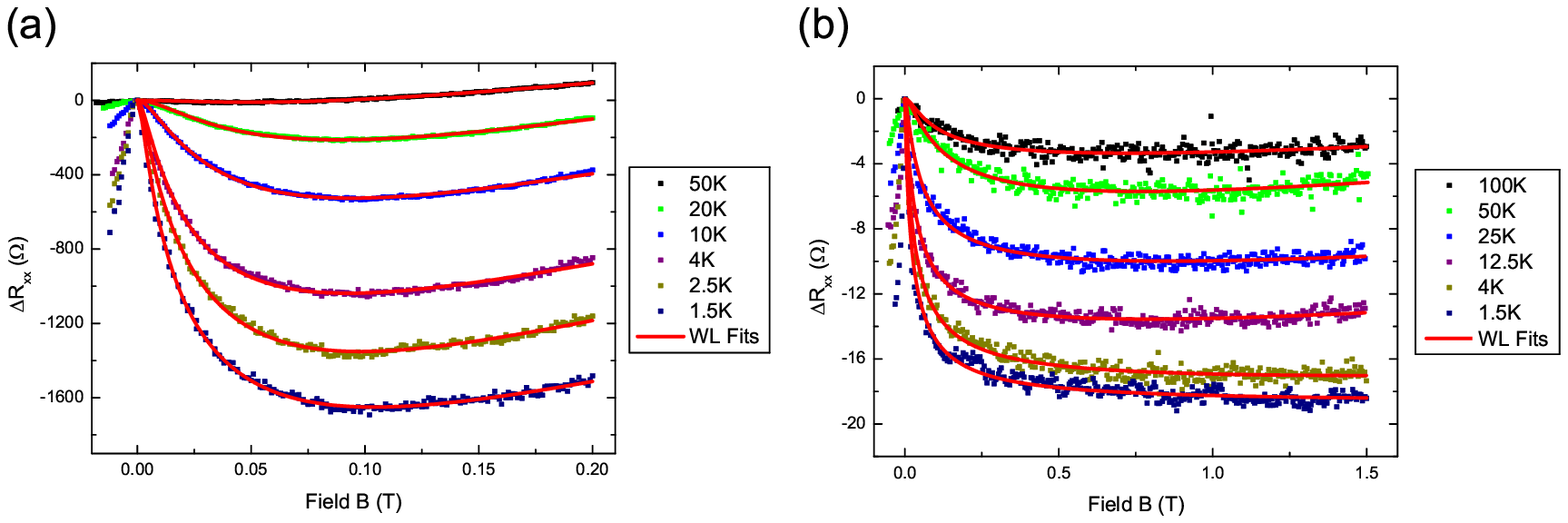}
\end{center}
\caption{Plots showing weak localization fits using the theory of McCann et al.\cite{McCann2006} from the two extremes of carrier density for the measured samples. The two plots highlight the dramatically different range in field and resistance that weak localization can occur over. (a) Epitaxially grown on SiC\cite{Janssen2011}, n = 1 x $10^{11}$\,cm$^{-2}$, exhibiting a low-temperature weak localization magnitude of  1.6\,k$\Omega$, and a minimum in R$_{xx}$ at 0.1\,T. (b) Grown by CVD onto copper\cite{Chen2012}, n = 1.43 x $10^{13}$\,cm$^{-2}$, exhibiting a low-temperature weak localization magnitude of  18\,$\Omega$, and a minimum in R$_{xx}$ in excess of 1.5\,T for low-temperatures.} \label{fig:fits}
\end{figure*} 

Weak (anti)localization is a quantum interference effect which occurs at low temperatures when electrons retain phase coherence\cite{Castro2009}.  Fig. \ref{fig:wltaus} shows four scattering terms which contribute to this process. Fig. \ref{fig:wltaus} (a) shows $\tau_\varphi$, the dephasing rate due to inelastic scattering\cite{McCann2006}. Fig. \ref{fig:wltaus} (b) shows the three other main scattering terms\cite{Tikh2008b}: $\tau_i$, the elastic intervalley scattering rate which comes from atomically sharp scatterers and scattering from the edges of the device, $\tau_w$, the elastic intravalley trigonal warping scattering term, and finally $\tau_z$, the elastic intravalley chirality breaking scattering term which comes from dislocations or other topological defects. These processes are grouped together as a single $\tau_*$ originally defined\cite{McCann2006} as $\tau^{-1}_* \equiv \tau^{-1}_w + \tau^{-1}_z + \tau^{-1}_i$. (The alternative definition of $\tau^{-1}_* = \tau^{-1}_w + \tau^{-1}_z$ is not used here).


Fig. \ref{fig:wltaus} (a) displays two self-intersecting scattering paths. These two paths are identical except for the direction of travel around the loop. Interference between such loops is the origin of the weak (anti)localization effect. If these paths constructively interfere, such loops are more common than would be expected classically, resulting in an increase in resistance known as weak localization. The converse, the destructive interference case, is called weak antilocalization. Due to the need to maintain phase coherence for an interference effect to occur, $\tau_\varphi$ acts to control the localization through the maximum size of such loops which is given by the decoherence length defined by $L_{\varphi} = \sqrt{\tau_{\varphi} D}$ where D, the diffusion coefficient = $\frac{1}{2} v_F^2\tau_{tr}$, $v_F$ is the Fermi velocity, which is 1.1 x 10$^6$ms$^{-1}$ as measured in both epitaxial SiC/G\cite{Miller2009} and exfoliated material \cite{Deacon2007} and $\tau_{tr}$ is the transport scattering time as determined from the carrier mobility. Hence $L_{\varphi}$ controls the magnitude of the weak (anti)localization effect.

Whether we are operating in a weak localization regime, or a weak antilocalization regime, depends on the phase the carriers pick up while traversing such a path. Because of the existence of a Berry phase in monolayer graphene\cite{Novoselov2005}, the two trajectories are expected to gain a phase difference of $\pi$, leading to destructive interference, and hence weak antilocalization\cite{Horsell2008}. However, in the presence of significant elastic intervalley scattering ($\tau_i$), weak localization can be restored. The reason for this is that chirality is reversed between the two valleys\cite{Falko2007}; hence trajectories involving intervalley scattering allow for zero phase difference between two self-intersecting paths which leads to constructive interference and hence weak localization.



The weak (anti)localization effect can be destroyed by increasing either the magnetic field or temperature to a sufficient value. Increased magnetic fields add a random relative phase to the carriers as they traverse curved paths, causing the interference effect to be averaged away\cite{Castro2009}. Increased temperature has the effect of decreasing $\tau_\varphi$, which reduces the magnitude of both types of localization effect, as can be seen from Eq. (\ref{eqn:mccann2}).

This paper makes use of the main result from McCann et al.\cite{McCann2006} to produce fits of the resistivity as a function of magnetic field, B, to the measured weak (anti)localization,

\begin{eqnarray}\label{eqn:mccann1}\nonumber
\Delta\rho(B)&=&-\frac{e^2\rho^2}{\pi h} \left(F\left(\frac{\tau^{-1}_B}{\tau^{-1}_\varphi}\right)-F\left(\frac{\tau^{-1}_B}{\tau^{-1}_\varphi+2\tau^{-1}_i}\right)\right. \\
&& \left.-2F\left(\frac{\tau^{-1}_B}{\tau^{-1}_\varphi+\tau^{-1}_*}\right)\right),
\end{eqnarray}

\noindent where $F(z) = \ln z + \psi \left(\frac{1}{2} + \frac{1}{z}\right)$, $\psi$ is the digamma function and $\tau^{-1}_B = \frac{4eDB}{\hbar}$. At small magnetic fields, where $z\ll1$, we can approximate $F(z) \approx \frac{z^2}{24}$. Using this we can simplify Eq. (\ref{eqn:mccann1}) for small fields as,


\begin{eqnarray}\label{eqn:mccann2}\nonumber
\Delta\rho(B)&=&-\frac{e^2\rho^2}{24\pi h} \left(\frac{4eDB\tau_\varphi}{\hbar}\right)^2  \Bigg(1-\frac{1}{\left(1+2\frac{\tau_\varphi}{\tau_i}\right)^2} \\ &&
-\frac{2}{\left(1+\frac{\tau_\varphi}{\tau_*}\right)^2}\Bigg).
\end{eqnarray}

\noindent From this equation it is clear how variations in $\tau_\varphi$ control the magnitude of the weak (anti)localization. It is also clear how significant intervalley scattering, $\tau_i$, is required to produce a positive resistivity correction, i.e. weak localization. In practice, significant intervalley scattering is found in most samples, and therefore, weak localization is far more commonly found than weak antilocalization\cite{Tikh2009}.


Fig. \ref{fig:fits} shows data from the extremes of carrier density of the measured samples. The samples are found to be very well fitted by the McCann theory\cite{McCann2006}, despite the two samples having very different magnitudes, shape, and field range for the localization. To attain the best possible fits care must be taken to avoid landing in local minima of the parameter space, especially when $\tau_*$ and/or $\tau_i$ are very short.

\section{Scattering Lengths}

\begin{figure}
\begin{center}
\includegraphics[width = 8.5cm]{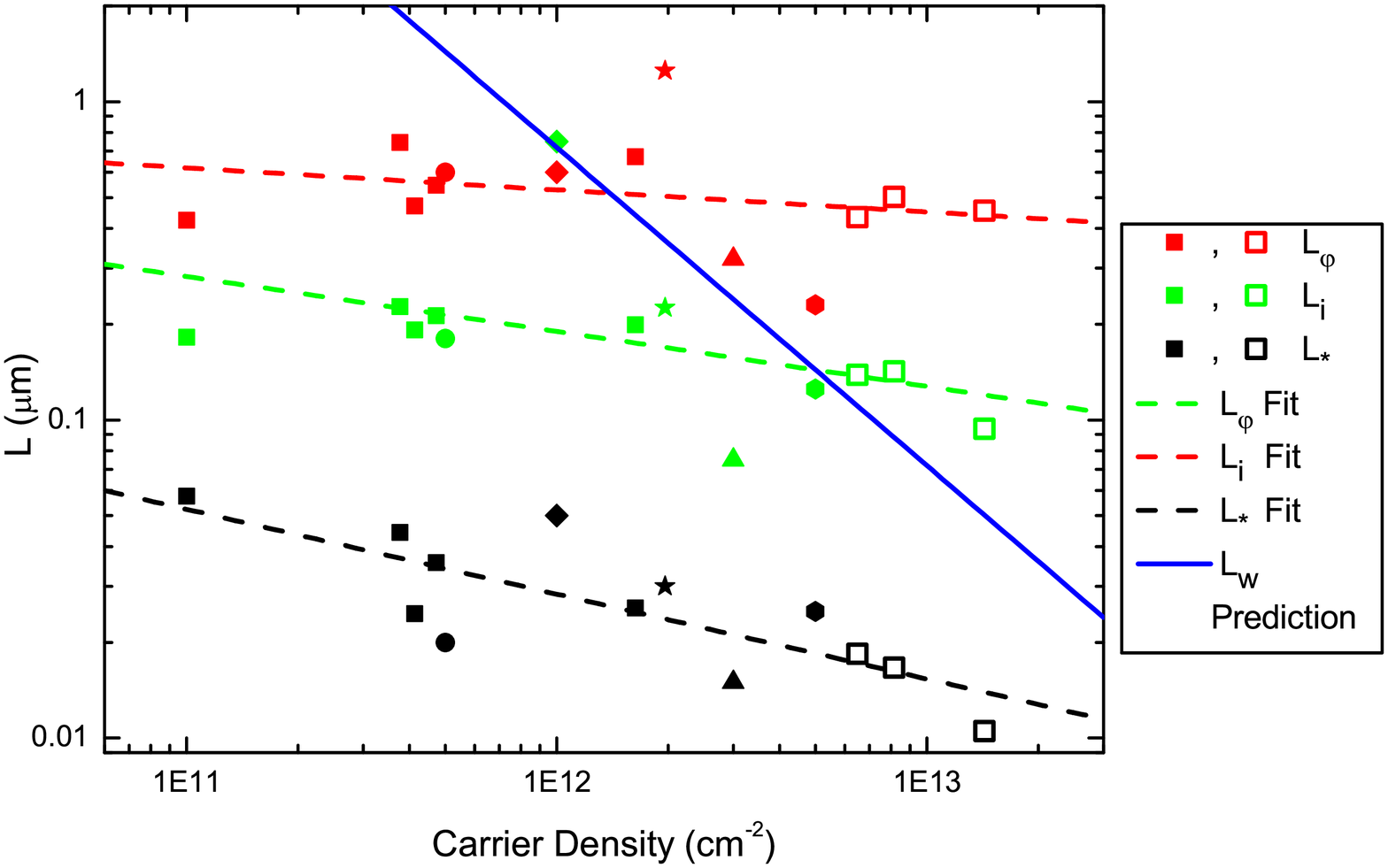}
\end{center}
\caption{Scattering lengths as a function of carrier density. Filled squares denote the data taken with the epitaxial devices and open squares CVD material, collected at 1.5\,K. Circles denote data from Lara-Avila et al.\cite{Lara2011b}. Diamonds from Tikhoneko et al.\cite{Tikh2008}. Stars from Ki et al.\cite{Ki2008}. Hexagons from Jauregui et al.\cite{Jaur2011}. Data collected from the literature are taken as close to 1.5\,K as possible. L$_w$ Prediction is from McCann et al.\cite{McCann2006}} \label{fig:allls}
\end{figure}

Fitting to the magnetoresistivity as shown in Fig. \ref{fig:fits} for 8 different samples with carrier densities from 1 x $10^{11}$\,cm$^{-2}$ to 1.4 x $10^{13}$\,cm$^{-2}$ allows us to extract the scattering times using Eq. (\ref{eqn:mccann1}) and these were converted to scattering lengths using, $L_{\varphi,i,*} = \sqrt{\tau_{\varphi,i,*} D}$. Fig. \ref{fig:allls} shows the extracted scattering lengths, from our data and from the literature\cite{Lara2011b,Tikh2008,Ki2008,Jaur2011}. Care was taken to extract all values for as close as possible to the same temperature, in this case 1.5\,K. This is done since L$_\varphi$ in particular is known to vary strongly with temperature\cite{Lara2011b,Tikh2008,Ki2008,Jaur2011}. Fits to the data are made using a simple power law, $Bn^A$, the results of which are shown in Table \ref{fig:Table}.

To within the standard error we find no variation with carrier density for the phase coherence length (L$_\varphi$) despite the very different physical nature of the epitaxial, exfoliated and CVD samples. Previous work has typically found similar values for L$_\varphi$ of around 0.6\,$\mu$m\cite{Lara2011b,Tikh2008,Ki2008,Jaur2011}. In Ki et al.\cite{Ki2008}, there has been some previous work carried out on the carrier density dependence by using a single sample with a backgate. In their work they found a superlinear increase of L$_\varphi$ with carrier density. These devices, however, were very small at 6 x 1 $\mu$m$^2$ and were probably effected by boundary scattering. More indirectly, temperature studies have also been carried out on L$_\varphi$, the modelling of which could in principle be used to predict a carrier density dependence. In Ki et al.\cite{Ki2008}, the behaviour of the scattering length is modelled using two electron-electron interaction terms, a direct Coulomb term and a Nyquist scattering term. These terms do have a carrier density dependence, however, the fitting parameters were found to vary with carrier density. Lara-Avila et al.\cite{Lara2011b} use an alternative model and find their data to be well modelled by the addition of a electron spin-flip scattering term. This is due to scattering from the localized magnetic moment of spin-carrying defects which is likely to be dependent on the sample preparation method and could mask or dominate underlying trends in the dependence of the phase coherence length on carrier density.

\begin{table}
\footnotesize
\begin{center}
\begin{tabular} {cccc}
\hline
Scattering& Exponent & Exponent& Multiplicative\\
Length&(A) & Standard Error&Constant (B) \\
\hline
L$_\varphi$& $-0.069$&$\pm 0.082$&3.59$\cdot 10^{-6}$\,m \\
L$_i$& $-0.173$ &$\pm 0.101$&2.25$\cdot 10^{-5}$\,m \\
L$_*$ & $-0.267$ &$\pm 0.064$&4.47$\cdot 10^{-5}$\,m \\
\hline
\end{tabular}
\end{center}
\caption{Multiplicative constant and exponents of the fits to the data in Fig. \ref{fig:allls} of the form Bn$^A$, where n is the carrier density in carriers per cm$^{2}$.}\label{fig:Table}
\end{table}

For the elastic intervalley scattering term, L$_i$, we find a weak trend with carrier density with a negative exponent of -0.173. Previous temperature\cite{Lara2011b,Ki2008} and backgate studies\cite{Ki2008} found no strong variation of L$_i$ with either temperature or carrier density. Given that L$_i$ is due to short range, atomically sharp scatterers and device-edge scattering, it would be expected to be highly dependent on the device characteristics. We might also expect that there would be some correlation with the the ungated carrier density as this is related to the number of defects through shifting of the Fermi level by the presence of charged defects\cite{Katoch2010}. In particular, for the data presented here, the highest ungated carrier densities are found for CVD graphene devices which are associated with high levels of polycrystallinity. This implies a large number of atomically sharp scatterers, and hence could account for the lower values of L$_i$ measured at high carrier densities using CVD samples.

The strongest trend, with an exponent of -0.267, and smallest standard error ($\pm0.064$) is found for $L_*$, the sum of all the sublattice-symmetry-breaking perturbations. For all samples in Fig. \ref{fig:allls}, L$_i$ $\gg$ L$_*$ and hence L$_*$ will be predominantly made up from L$_w$, the elastic intravalley trigonal warping scattering term, and L$_z$, which allows for other chirality breaking elastic intravalley processes. We would expect the trigonal warping term to increase with carrier density, since the degree of trigonal warping is dependent on the Fermi energy\cite{McCann2006}. The L$_z$ term is expected to be relatively independent of carrier density due to its origin from topological defects\cite{Tikh2008b}. McCann et al.\cite{McCann2006} produce the following prediction of how $L_w$ is expected to vary with carrier density,

\begin{equation}\label{eqn:lw1}
L_w^{-2} = \frac{\tau_w^{-1}}{D} = \frac{\tau_{tr}}{D}\left(\frac{\mu E_F^2}{\hbar v_F^2}\right)^2 \propto n^2.
\end{equation}

\noindent where $\mu$, the structure constant $= \frac{\gamma_0a^2}{8\hbar^2}$, $\gamma_0$ is the nearest neighbour overlap integral, $a$ is the lattice constant, E$_F$ is the Fermi energy, and $v_F$ is the Fermi velocity. This equation predicts that L$_w$ should be proportional to $\frac{1}{n}$, and is shown in Fig. \ref{fig:allls} as a solid blue line suggesting that the trigonal warping term will not become dominant until around 1\,\,x\,\,$10^{14}$\,cm$^{-2}$. For the region studied we find a much slower variation with n of approx n$^{\frac{1}{4}}$ suggesting that L$_z$ is dominant and only varies weakly with carrier density. 


\begin{figure} [htb]
\begin{center}
\includegraphics[width = 7cm]{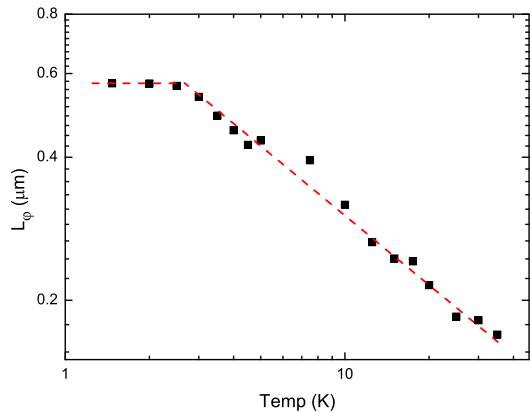}
\end{center}
\caption{L$_\varphi$ data for for an epitaxial sample with a carrier density of 4.72 x $10^{11}$\,cm$^{-2}$, a size of 160 x 35\,$\mu$m$^2$, and a sample resistance at zero field of 8.2\,k$\Omega$. All data was measured with a current of 500\,nA.} \label{fig:samplecurrentcap}
\end{figure}

\section{Maximum Currents}

In this section the importance of using sufficiently low currents is demonstrated, together with how the use of too large currents may explain the observations of a ``saturation'' in L$_\varphi$ sometimes found in the literature. Because of the very large optical phonon energies in graphene\cite{Yan2007}, the dominant cooling mechanism for carriers at low temperatures comes from acoustic phonons\cite{Baker2012}. The acoustic phonon cooling in graphene is a fairly weak mechanism which allows carriers to attain temperatures far in excess of that of the lattice\cite{Baker2012,Baker2012b}, and at low temperatures in the Bloch-Gr\"{u}neisen limit this process is strongly temperature dependent. This ``hot-carrier'' effect can be described using the theory of Kubakaddi\cite{Kubakaddi2009}, which has been shown experimentally to predict the energy loss rates very accurately\cite{Baker2012,Baker2012b}. Using this theory, we can calculate the effective minimum carrier temperature, T$_{e,min}$, that can be obtained for a given device for each current. Kubakaddi presents the relation for the energy loss rate per carrier,

\begin{equation}\label{eqn:elr1}
F(T)  = \alpha(T_e^4-T_l^4),
\end{equation}

\noindent where T$_e$ is the carrier temperature, and T$_L$ is the lattice temperature. For a given current and sample resistance R$_{xx}$, this can be equated to the power input per carrier from the current as

\begin{equation}\label{eqn:elr2}
\alpha(T_e^4-T_l^4) = \frac{I^2 R_{xx}}{nA},
\end{equation}

\noindent where n is the carrier density, and A is the sample area. The coefficient $\alpha$ is calculated using the relation

\begin{equation}\label{eqn:elr3}
\alpha = \frac{D^2E_Fk_B^43!\zeta(4)}{n\pi^2\rho\hbar^5v_s^3v_f^3},
\end{equation}

\noindent where $\zeta$ is the Riemann zeta function, $\rho$ is the sample density, and $v_s$ is the sound velocity. This can be rearranged to give the effective minimum carrier temperature,

\begin{equation}\label{eqn:elr4}
T_{e,min} = \sqrt[4]{\frac{I^2R_{xx}}{\alpha n A}+T_L^4},
\end{equation}

\noindent  Using the numerical values suggested by Kubakaddi\cite{Kubakaddi2009} we calculate $\alpha$ = 5.36  x $10^{-18}$\,W K$^{-4}$ / $\sqrt{n}$, where n is in units of \,10$^{12}$\,cm$^{-2}$.

Fig. \ref{fig:samplecurrentcap} shows data from one of our epitaxial samples which exhibits a saturation in the measured value of L$_\varphi$ with decreasing temperature. The sample has a carrier density of 4.72 x $10^{11}$\,cm$^{-2}$, a size of 160 x 35\,$\mu$m$^2$ and a sample resistance of 8.2\,k$\Omega$. All the data for the graph was collected with a current of 500\,nA. Using Eq. (\ref{eqn:elr4}) we calculate T$_{e,min}$ for the sample as 1.97\,K. This value corresponds well to the temperature of the measured onset of the saturation regime presented in the figure.

\begin{figure} [htb]
\begin{center}
\includegraphics[width = 8.5cm]{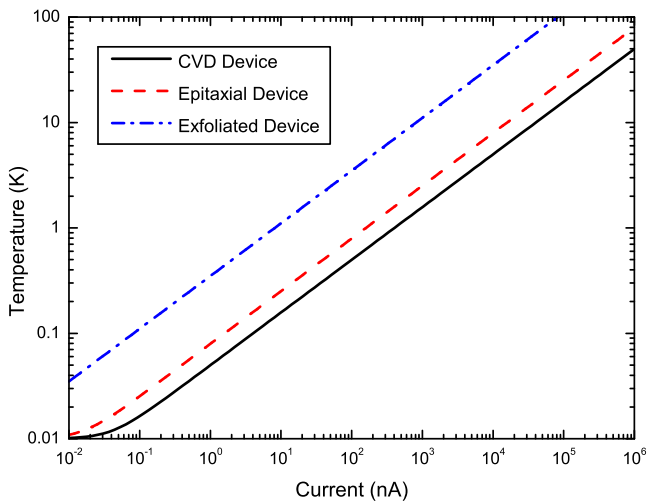}
\end{center}
\caption{The minimum carrier temperatures obtainable for a given current, for three example devices, calculated for a lattice temperature, T$_L$ of 10\,mK. The epitaxial device is as used in Fig. \ref{fig:samplecurrentcap} (n = 4.72 x $10^{11}$\,cm$^{-2}$, A = 160 x 35\,$\mu$m$^2$, R$_{xx}$ = 8.2\,k$\Omega$). The CVD device is as used in Fig. \ref{fig:fits} (b) (n = 1.43 x $10^{13}$\,cm$^{-2}$, A = 64 x 16\,$\mu$m$^2$, R$_{xx}$ = 1.3\,k$\Omega$). The final exfoliated device is for a typical small device as commonly used in the literature (n = 1 x $10^{12}$\,cm$^{-2}$, A = 5 x 1\,$\mu$m$^2$, R$_{xx}$ = 4\,k$\Omega$).} \label{fig:allcurrentcap}
\end{figure} 

When previously encountered, this saturation in measured L$_\varphi$ at quite high temperatures has been attributed variously to magnetic impurities\cite{Ki2008}, electron-hole puddles reducing the effective conducting area\cite{Ki2008}, and limits imposed directly from the sample size\cite{Tikh2008}. We believe the above hot-carrier effects should also be taken into account, particularly when the sample size is physically small. Giving further weight to the validity of the hot-carrier explanation, Lara-Avila et al.\cite{Lara2011b} showed that significant changes in L$_\varphi$ could still be observed at temperatures below 100\,mK by using a large area device and a current of 50\,pA for which Eq. (\ref{eqn:elr4}) predicts T$_e$ $\sim$ 20\,mK. Further evidence for electron temperature saturation in graphene has been observed recently through measurements of bolometric response in noise power\cite{Fong2012},\cite{Betz:2012} and resistivity \cite{Yan2012} which require similar energy loss rates to those used here\cite{Kubakaddi2009,Baker2012,Baker2012b}.

By way of illustration we calculate the currents required to achieve a given carrier temperature for three different examples of typical samples used here and in the literature which we present in Fig. \ref{fig:allcurrentcap}. The epitaxial device in the figure is the one from Fig. \ref{fig:samplecurrentcap}, the CVD device is the one from Fig. \ref{fig:fits} (b), and the third, exfoliated graphene device is a typical device similar to many of those used in the literature of dimensions 5 x 1\,$\mu$m$^2$. It is striking, and worth emphasizing, that for this device to attain a carrier temperature of 30\,mK, it requires maximum currents of $\sim$0.01\,nA.

\section{Conclusions}

Using the theory of McCann et al.\cite{McCann2006} we have shown that high quality fits to weak-localization can be obtained for devices with carrier densities from 1 x $10^{11}$\,cm$^{-2}$ to 1.43 x $10^{13}$\,cm$^{-2}$ for graphene fabricated by both the epitaxial and CVD methods.  We have investigated carrier density dependences for L$_\varphi$, L$_i$, and L$_*$. We find no evidence of a significant density dependence for L$_\varphi$ and only a weak decrease in L$_i$ with increasing density, though this may be due to a coincidental increase in disorder. Finally, we find evidence of a weak power law decrease in L$_*$ with a carrier density dependence of approximately n$^{-\frac{1}{4}}$. We have also shown that hot electron effects may obscure the true temperature dependence of the scattering lengths unless currents as low as 0.01\,nA are used for measurements at dilution fridge temperatures in small devices.

\section{Acknowledgements}
This work was supported by the UK EPSRC, Swedish Research Council and Foundation for Strategic Research, UK National Measurement Office, and EU FP7 STREP ConceptGraphene.




\end{document}